\documentclass[11pt,showpacs]{revtex4}
\input epsf.sty

\usepackage{ulem}
\usepackage{cancel}
\usepackage{color}

\def\comment#1{}

\def\E{{\mathcal E}}

\usepackage{graphicx}
\begin{document}
\title{Higgs Boson and Top-Quark Masses and Parity-Symmetry Restoration}
\author{She-Sheng Xue}
\email{xue@icra.it}
\affiliation{ICRANeT, Piazzale della Repubblica, 10-65122, Pescara,\\
Physics Department, University of Rome ``La Sapienza'', Rome,
Italy} 


\begin{abstract}
The recent ATLAS and CMS experiments show the first observations of a new particle in the search for the Standard Model Higgs boson at the LHC. We revisit the scenario that high-dimensional operators of fermions must be present due to the theoretical inconsistency of the fundamental cutoff (quantum gravity) with the parity-violating gauge symmetry of the Standard Model. 
Studying the four-fermion interaction of the third quark family, we show that at an intermediate energy threshold $\E\approx 4.27\times 10^3\,$GeV for the four-fermion coupling being larger than a critical value, the spontaneous symmetry-breaking phase transits to the strong-coupling symmetric phase where composite Dirac fermions form fully preserving the chiral gauge symmetry of the Standard Model and the parity-symmetry is restored. Under this circumstance,
we perform the standard analysis of renormalization-group equations of the Standard Model in the spontaneous symmetry-breaking phase. As a result, the Higss-boson mass $m_H\approx 126.7$GeV and top-quark mass $m_t\approx 172.7$GeV are obtained without drastically fine-tuning the four-fermion coupling.   
\end{abstract}


\maketitle

\noindent
{\bf Introduction.}
\hskip0.1cm
Since its appearance, the Standard Model for particle physics has always been extremely peculiar. The parity-violating (chiral) gauge couplings and spontaneous/explicit breakings of these symmetries for the hierarchy of fermion masses have been at the center of a conceptual elaboration and 
an intensive experimental analysis that have played a major role in donating to mankind the beauty of the Standard Model for particle physics. 
The Nambu-Jona-Lasinio model \cite{njl} for high energies and its effective counterpart for low energies, the Higgs model \cite{higgs}, provide an elegant description for the
electroweak breaking scale, intermediate gauge boson masses and their relations. After a great experimental effort for many years, the ATLAS \cite{ATLAS} and CMS \cite{CMS} experiments have recently shown the first observations of a 126 GeV scalar particle in the search for the Standard
Model Higgs boson at the LHC. This far-reaching result
begins to shed light on this most elusive and fascinating arena of fundamental particle physics. 

When the top quark mass $m_t$ was discovered to be greater than $\sim 10^2\,$GeV, several authors \cite{nambu1989,Marciano1989,mty1989,bhl1990} in 1989 suggested that the symmetry breakdown of the Standard Model could be a dynamical mechanism of the Nambu-Jona-Lasinio or BCS type that intimately involves the top quark at a high-energy scale $\Lambda$. This dynamical mechanism leads to the formation of a low-energy $\bar t t$-condensate, which is responsible for the top quark, $W^\pm$ and $Z^\circ$ gauge bosons masses, and a composite particle of the Higgs type.
Since then, many models based on this idea have been proposed and studied \cite{DSB_review}. 
For our following discussions, we will adopt the model for the minimal dynamical symmetry breaking via an effective four-fermion operator of the Nambu-Jona-Lasinio type
\begin{eqnarray}
L = L_{\rm kinetic} + G(\bar\psi^{ia}_Lt_{Ra})(\bar t^b_{R}\psi_{Lib}),
\label{bhl}
\end{eqnarray}
which was studied by Bardeen, Hill and Lindner (BHL) \cite{bhl1990} in the context of a well-defined quantum field theory at the high-energy scale $\Lambda$.  The fermion fields in $L_{\rm kinetic}$ are massless, and the
four-fermion coupling $G\sim 1/\Lambda^2$.

To achieve the low-energy electroweak scale for the top quark mass $m_t$ by the renormalization group equations \cite{bhl1990,Marciano1989,bhl1990a}, this model (\ref{bhl}) requires $\Lambda/m_t \gg 1$ with a drastically unnatural fine tuning, which is known as the gauge hierarchy problem, and the top quark mass $m_t$ is determined by the infrared quasi-fixed point \cite{bhl1990a}. To have a natural scheme incorporating the effective four-fermion operator of the Nambu-Jona-Lasinio type (\ref{bhl}), some strong technicolor dynamics at the TeV scale were invoked \cite{hill1994}. This scheme is preferentially coupled to the third quark family of top and bottom quarks. The possibility of the $126$ GeV particle being a light pseudoscalar, such as the top-pion \cite{bhl1990a}, seems unlikely because the loop-suppressed couplings of light pseudoscalars to the Standard Model gauge bosons are too small to generate the observed signal \cite{top-pion}.
These discussions indicate that much effort is still required to study the issue of the minimal dynamical symmetry breaking that is preferentially associated with the top quark (the top-Higgs system) in the theoretical aspects of dynamics or/and symmetry (see for example \cite{Hill2013}) to discover if the issue agrees with experiments.  

Suppose that the effective high-dimensional
operators of all fermion fields, for example Eq.~(\ref{bhl}), are generated by the new dynamics at the scale $\Lambda$, which will be discussed in the end of the Letter. It is conceivable that the new dynamics at the scale $\Lambda$ should be on an equal footing with all the fermions in the Standard
Model because the scale $\Lambda$ is much larger than the masses of
all the fermions. This raises a neutral question: why should
the new dynamics preferentially act on the top-quark alone?
In our recent Letter \cite{xue2013}, we understand, from the dynamical point of view, a compelling possible answer to this question by studying the following effective four-fermion operator: 
\begin{eqnarray}
L &=& L_{\rm kinetic} + G(\bar\psi^{ia}_L\psi_{Rja})(\bar \psi^{jb}_R\psi_{Lib}),\nonumber\\
&=& L_{\rm kinetic} + G(\bar\psi^{ia}_Lt_{Ra})(\bar t^b_{R}\psi_{Lib})
+ G(\bar\psi^{ia}_Lb_{Ra})(\bar b^b_{R}\psi_{Lib}),
\label{bhlx}
\end{eqnarray}
where $a,b$ and $i,j$ are, respectively, the color and flavor indexes of the top and bottom quarks, the left-handed doublet $\psi_L=(t_L,b_L)$ and the right-handed singlet $\psi_R=t_R,b_R$. By calculating the vacuum energy of Eq.~(\ref{bhlx}) we show that the minimal dynamical symmetry
breaking Eq.~(\ref{bhl}) for the top-quark is an energetically favorable configuration (the
ground state) of the quantum field theory with the high-dimension operators of all the fermion fields at the cutoff $\Lambda$. This result is not surprising. One can see that the vacuum energy decreases (the system of fields gains energy) as the fermions acquire their masses by the spontaneous chiral symmetry breaking; however, the associated scalar and pseudoscalar modes have positive contributions to the vacuum energy. 
Three pseudoscalar (Goldstone) modes become the longitudinal modes of the intermediate gauge bosons $W_\mu^\pm$ and $Z^\circ$. As more fermions acquire their masses by the spontaneous chiral symmetry breaking, more associated scalar and pseudoscalar modes are produced. As a result, the energetically favorable configuration is the one in which only one quark (the top quark) acquires its mass by the spontaneous chiral symmetry breaking, with three pseudoscalar modes as the longitudinal modes of the massive gauge bosons and a scalar particle of the Higgs type. 
In addition, 
we discussed the strong-coupling symmetric phase where composite Dirac fermions form and the vector-like feature of $W^\pm$-boson coupling, which leads to the explicit symmetry breaking for generating masses of other fermions.

In this Letter, we present the study of strong four-fermion interaction of the third quark family, and show that at an intermediate energy threshold $\E\approx 4.27\times 10^3\,$GeV for the four-fermion coupling being larger than a critical value $G_{\rm crit}=2N_c(\pi/\Lambda)^2$, the spontaneous symmetry-breaking phase transits to the strong-coupling symmetric phase where composite Dirac fermions form fully preserving the chiral gauge symmetry of the Standard Model and the parity-symmetry is restored. Taking duly into account this phase transition,
we perform the standard analysis of renormalization-group equations of the Standard Model in the spontaneous symmetry-breaking phase. As a result, the Higgs boson mass $m_H\approx 126.7$GeV and top-quark mass $m_t\approx 172.7$GeV are obtained without drastically fine-tuning the four-fermion coupling.
The natural units $\hbar=c=1$ are adopted, unless otherwise specified. 

\noindent
{\bf The weak-coupling phases.}
\hskip0.1cm 
Employ the ``large $N_c$-expansion'' for weak coupling $G$, i.e., keep $GN_c$ fixed and construct the theory systematically in powers of $1/N_c$. At the lowest order, one has the gap equation for the induced top-quark masses $m_t=-G\langle \bar tt\rangle$:
\begin{eqnarray}
m_t&=& 2GN_c\frac{i}{(2\pi)^4}\int_\Lambda d^4l(l^2-m^2)^{-1}m_t.
\label{gap0}
\end{eqnarray}
In addition to the trivial solution $m_t=0$, the gap equation (\ref{gap0}) has a non-trivial solution $m_t\not=0$ 
\begin{eqnarray}
\frac{1}{G_c}-\frac{1}{G}=\frac{1}{G_c}\left(\frac{m_t}{\Lambda}\right)^2
\ln \left(\frac{\Lambda}{m_t}\right)^2>0,
\label{delta}
\end{eqnarray}
when the coupling $G\geq G_c\equiv 8\pi^2/(N_c\Lambda^2)$, where $G_c$ is the ``critical'' weak-coupling constant.    The theory (\ref{bhl}) is in the weak-coupling symmetric phase $m_t=0$ for $G<G_c$, or in the symmetry-breaking phase $m_t\not=0$ for $G> G_c$. The result (\ref{delta}) is the leading order of large $N_c$-expansion, it becomes exact in the weak-coupling limit: $GN_c\rightarrow {\rm finite}$ when $N_c\rightarrow \infty$. 
Eq.~(\ref{delta}) needs a drastically fine-tuning $G=G(\Lambda, m_t)\rightarrow G_c$ for $m_t\ll \Lambda$.  
 
\noindent
{\bf The strong-coupling symmetric phase.}
\hskip0.3cm
In the strong-coupling limit $Ga^{-2}\gg 1$, where we introduce the lattice spacing $a\equiv (\pi/\Lambda)$, the theory (\ref{bhl}) is in the strong-coupling symmetric phase (see Refs.~\cite{ep1986,xue1997}). Using the Lagrangian (\ref{bhlx}), we briefly review the strong-coupling symmetric phase based on Ref.~\cite{xue1997}. In order to perform the strong coupling expansion in powers of $1/g$, we rescaled all fermion fields to dimensionless fields,
\begin{equation}
\psi(x)\rightarrow \psi(x)=a G^{1/4}\psi(x) =a^2 g^{1/4}\psi(x),\quad g\equiv G/a^4
\label{rescale}
\end{equation}
and rewritten the fermion action in terms of the dimensionless fields on the lattice
\begin{eqnarray}
S_{\rm kinetic}&=&{1\over 2ag^{1/2}}\sum_{x,\mu}\bar\psi(x)
\gamma_\mu \partial^\mu\psi(x),\quad \partial^\mu\equiv \delta_{x,x+a_\mu}-\delta_{x,x-a_\mu}\label{rfa}\\
S_{\rm int}&=&\sum_{x}\left[(\bar\psi^{ia}_Lt_{Ra})(\bar t^b_{R}\psi_{Lib}) + (\bar\psi^{ia}_Lb_{Ra})(\bar b^b_{R}\psi_{Lib})\right].\label{rs2}
\end{eqnarray}
where all weak gauge couplings are neglected. 
In the strong coupling limit $ga^2\gg 1$, treating the kinetic
action $S_{\rm kinetic}$ as a small perturbation, we calculated two-point function of fermion fields by the strong coupling (hopping) expansion in powers of ${1/g}$. As a result, in the lowest non-trivial
order we obtained the propagators ($p^\mu a< 1$)
\begin{equation}
S_F(p)\simeq {ip^\mu\gamma_\mu+M\over
p^2+M^2},
\label{sc1}
\end{equation}
of the composite massive Dirac fermions: $SU_L(2)$-doublet ${\bf\Psi}^{ib}_D=(t^{ib}_L,{\bf\Psi}^{ib}_R)$ and $SU_L(2)$-singlet ${\bf\Psi}^{b}_D=({\bf\Psi}^b_{L},t_R^b)$, where the composite three-fermion states are:
\begin{equation}
{\bf\Psi}^{ib}_R=[Z^{^S}_F]^{1/2}{g\over
2a}(\bar\psi^{ia}_Lt_{Ra})t^b_{R},\quad {\bf\Psi}^b_{L}=[Z^{^S}_F]^{1/2}{g\over
2a}(\bar\psi^{ia}_Lt_{Ra})\psi^{b}_{iL},
\label{bound}
\end{equation}
$[Z^{^S}_F]$ and $M$ are respectively the form-factor (wave-function renormalization) and mass of composite Dirac fermions. We need to stress that the composite Dirac fermion propagator (\ref{sc1}), $Z^{^S}_F=1$ and $ M=2ga /Z^{^S}_F$ are obtained by considering $S_{\rm int}$ of Eq.~(\ref{rs2}) and only one ``hopping'' step ($1/g^{1/2}$) of Eq.~(\ref{rfa}) at the cutoff scale $a=\pi/\Lambda$. It is difficult to do the calculations of many ``hopping'' steps to obtain the energy-momentum dependence of $Z^{^S}_F(p)$ and $M(p)$ down to some scales $\mu$ smaller than the cutoff $\Lambda$.

In the strong-coupling symmetric phase, the three-fermion state ${\bf\Psi}^{ib}_R$ (${\bf\Psi}^{b}_L$) is the bound state of a composite boson $H^i=(\bar\psi^{ia}_Lt_{Ra})$ and a fermion $t^b_{R}$ ($\psi^{b}_{iL}$). 
The discussions are the same for the bottom quark $t_{Ra}\rightarrow b_{Ra}$. 
These three-fermion states (\ref{bound}) carry the appropriate quantum
numbers of the chiral gauge group of the Standard Model that accommodates $\psi^{ia}_{L}$ and  $t_{Ra}$. Therefore 
massive composite Dirac fermions are consistent with the chiral symmetry $SU_L(2)\otimes SU_R(2)$ \cite{neutral}, and their couplings to intermediate gauge bosons are vector-like \cite{xue_parity,r-neutrino}, for example  
\begin{eqnarray}
\Gamma^{ij}_\mu (p,p')&=&i\frac{g_2}{2\sqrt{2}}U_{ij}\gamma_\mu f(p,p'),\quad q=p'-p
\label{wv}
\end{eqnarray}
where $g_2$ is the $SU_L(2)$ coupling, $U_{ij}$ the CKM matrix, $p,p'$ and $q$ are respectively composite fermion and gauge boson momenta. The vector-like form factor $f(p,p')$ of chiral-gauge coupling (\ref{wv}) is related to the chiral-symmetric mass $M(p)$ in Eq.~(\ref{sc1}) by the Ward identity of chiral gauge symmetries. Consequently the parity-symmetry is conserved in this strong-coupling symmetric phase \cite{xue1997,xue_parity}.

In order to determine the critical value $g_{\rm crit}$ that separates the strong-coupling symmetric phase 
from the symmetry-breaking phase ($m_t\not=0$),  we calculated the two-point functions of composite boson fields $H^i$ by the strong coupling (hopping) expansion in powers of ${1/g}$ \cite{xue1997}. As a result, in the lowest non-trivial order we obtained the propagator of massive composite bosons $H^i$ $(q^\mu a< 1)$,
\begin{eqnarray}
S^{ij}_B(q)\simeq [Z^{^S}_H]^{-1}{\delta_{ij}\over q^2
+\mu^2_H},\quad \mu^2_H= {4\over N_c}\left(g-{2N_c\over a^2}\right)[Z^{^S}_H]^{-1},
\label{mas}
\end{eqnarray}
where $[Z^{^S}_H]^{1/2}=1$ and $\mu_H$ respectively are the form-factor and mass of composite bosons.
Thus,
$\mu^2_HHH^{\dagger}$
gives the mass term of the composite bosons $H$ in
the effective Lagrangian. 

In the lowest non-trivial order of the strong-coupling expansion, the contribution to the 1PI vertex-coupling $\lambda_0$ of the self-interacting term $(HH^{\dagger })^2$ is suppressed by $(1/g)^2$. The 1PI vertex-coupling $\lambda_0\gtrsim 0$ is small, but positive, 
the energy
of ground states of the theory is bound from the bellow. The
mass term $\mu^2_HHH^{\dagger}$ changes sign from $\mu^2_H>0$ to $\mu^2_H<0$, indicating a spontaneous symmetry breaking $SU(2)\rightarrow U(1)$ occurs, and non-zero vacuum expectational value ($v\not=0$) is developed.
Eq.~(\ref{mas}) for $\mu^2_H=0$ gives rise to the critical strong-coupling $G_{\rm crit}$: 
\begin{equation}
g_{\rm crit}=2N_c/a^2,\quad G_{\rm crit}=2N_c a^2 > G_c
\label{gc}
\end{equation}
where the second order phase transition from the strong-coupling symmetric phase 
to the symmetry-breaking phase takes place. Note that the inequality is valid, considering that $G_c$ should be calculated for $N_c\gg 1$. 

Because we are not able to obtain the energy-momentum dependence of form-factors and masses of composite fermions and bosons, and other 1PI-functions of high-dimensional operators, as well as their renormalization group equations in the neighborhood of the second order phase transition, therefore 
we cannot give a detailed description of the dynamics occurring at the phase transition. However, 
we conceive that fermion energy-momenta $(p,p')$ and energy transfer $(q)$ decrease down to the certain energy threshold $\E$, the effective interacting vertex $\Gamma^{(4)}(p,p',q)$ of Eq.~(\ref{bhlx}) becomes small enough that the binding energies $E_{\rm bind}[G(a),a]$ of the three-fermion bound states (\ref{bound}) vanish. As a result, these bound states (\ref{bound}) dissolve into their constituents \cite{pole-cut}, the mass term $M(p)$ vanishes and the vector-like form factor $f(p,p')\rightarrow P_L=(1-\gamma_5)/2$. This restores the chiral-gauged fermion spectra and couplings, as described by the Standard Model \cite{c-symmetry}.  
We postulated \cite{xue1997,xue_parity} that the  energy threshold ${\mathcal E}$ is in the range 
\begin{equation}
v\ll \E < \Lambda,
\label{inter}
\end{equation}
where $v$ is the electroweak breaking scale.  Numerical non-perturbative calculations are required to verify the postulation.

\noindent
{\bf Renormalization-group boundary condition at high energies.}
\hskip0.1cm 
First let us consider the four-fermion coupling $G$ of the quantum field theory (\ref{bhl}) defined at the high-energy scale $\Lambda$ is smaller than the critical value $G_{\rm crit}$, i.e. $G< G_{\rm crit}$.  
Therefore the theory is in the symmetry-breaking phase, contains the spectra of fundamental fermions 
$\psi$ and composite bosons $H$. Following the prescription of Ref. \cite{bhl1990}, 
at a renomalization scale $\mu$ below the scale $\Lambda$ the effective Lagrangian of the theory is written as
\begin{equation}
L=L_{\rm kinetic} + g_{t0}(\bar \Psi_L t_RH+ {\rm h.c.}) + Z_H|D_\mu H|^2-m_{_H}^2H^\dagger H
-\frac{\lambda_0}{2}(H^\dagger H)^2.
\label{eff}
\end{equation} 
where $g^2_{t0}(\Lambda)/m^2_H(\Lambda)=G$ and $m_H(\Lambda)=\Lambda$. The conventional renormalization $Z_\psi=1$ for fundamental fermions and the unconventional wave-renormalization $Z_H \not =1$ for composite Higgs bosons $H$ are adopted. Thus the coupling constants, such as $\bar g_t$ and $\bar\lambda$ are renormalized at the scale $\mu$
\begin{equation}
\bar g_t(\mu)=\frac{Z_{HY}}{Z_H^{1/2}}g_{t0}, \quad \bar\lambda(\mu)=\frac{Z_{4H}}{Z_H^2}\lambda_0,
\label{renc}
\end{equation}
where $Z_{HY}$ and $Z_{4H}$ are proper renormalization constants of the Yukawa-coupling and quartic vertex in Eq.~(\ref{eff}). The Higgs field $H$ is dynamical with a vanishing wave-function renormalization constant at the scale  $\Lambda$, leading to the following boundary 
conditions
\begin{equation}
Z_{H}(\Lambda)=0, \quad\lambda_0(\Lambda)=0, \quad \bar g^2_t(\Lambda)=g^2_{t0}=G\Lambda^2
={\rm const} < G_{\rm crit}\Lambda^2.
\label{boun1}
\end{equation}
In Eq.~(\ref{eff}), the transformation $H\rightarrow H/\bar g_t(\mu)$ transforms the conventional normalization into the that required by Eq.~(\ref{boun1}), one thus has
\begin{equation}
\tilde Z_{H}(\mu)=1/\bar g^2_t(\mu), \quad \tilde \lambda(\mu)=\bar\lambda(\mu)/\bar g^4_t(\mu),
\label{boun0}
\end{equation}
where the tilde will henceforth denote the normalization convention appropriate for compositeness.

We turn to the situation that the strong-coupling symmetric phase appears 
for $G> G_{\rm crit}$. In this phase, according to the massive spectra (\ref{sc1},\ref{mas}) of composite Dirac fermions and bosons we obtained, the effective Lagrangian can be written as ($\mu^2_H>0$)
\begin{eqnarray}
L &=&Z^{^S}_F\bar{\bf\Psi}^{ib}_D(i\gamma_\mu D_\mu - M){\bf\Psi}^{ib}_D + Z_F\bar{\bf\Psi}^{b}_D(i\gamma_\mu D_\mu - M){\bf\Psi}^{b}_D \nonumber\\
&+& Z^{^S}_H|D_\mu H|^2-\mu^2_H H^\dagger H
-\frac{\lambda^{^S}_0}{2}(H^\dagger H)^2+\cdot\cdot\cdot,
\label{eff2}
\end{eqnarray}
at the scale $\mu$ ($\E<\mu<\Lambda$) being smaller than the scale $\Lambda$ but larger than an intermediate energy scale $\E$ of Eq.~(\ref{inter}), which implies the energy scale of the second order phase transition $G(\E)=G_{\rm crit}$
from the strong-coupling symmetric phase to the symmetry-breaking phase (\ref{eff2}). 
The wave-function renormalization constant of composite Dirac fermions (\ref{bound}) is 
$Z^{^S}_F=Z^{^S}_H\cdot Z_\psi=Z^{^S}_H$. 
Note that we use normal fermion and boson fields in Eq.~(\ref{eff2}), which are not dimensionless fields (\ref{rescale}). The composite boson mass $\mu_H$ in the symmetric phase is different from the one $m_{_H}$ in Eq.~(\ref{eff2}), the latter is the Higgs boson mass relating to the scale of spontaneous symmetry-braking. The wave-function renormalization constant 
$Z^{^S}_H$ (inverse form-factor) of composite bosons is also different from the one $Z_H$ of Eq.~(\ref{eff}), however, $Z^{^S}_H$ and $Z_H$ should match each other at 
the energy scale $\E$ and critical point $G(\E)=G_{\rm crit}$, where the phase transition occurs. The same discussion is applied to the 1PI quartic vertex-coupling $\lambda^{^S}_0$ and $\lambda_0$.

At the energy threshold $\E$, we approximately treat $Z^{^S}_H\not=0$ and $\lambda^{^S}_0\gtrsim 0$ as parameters, because we cannot calculate their evolutions with the scale $\mu$ from $\Lambda$ to $\E$. 
The boundary condition (\ref{boun1}) at the scale $\Lambda$ should be modified into the following boundary condition at the scale  $\E$, 
\begin{equation}
Z_{H}(\E)=Z^{^S}_{H}(\E)={\rm const}, \quad\lambda_0(\E)=\lambda^{^S}_0(\E)\gtrsim 0, \quad 
\bar g^2_t(\E)=g^2_{t0}=G_{\rm crit}\E^2.
\label{boun2}
\end{equation}
Because the wave-function renormalization $Z_{H}(\E)$ does not vanish at the energy scale $\E$, the coupling $\bar g_t(\E)$ of Eq.~(\ref{renc}) does not go to infinity. Based on the definitions (\ref{boun0}), this leads to the boundary condition $\tilde Z_{H}(\E)\not=0$ and 
$\tilde \lambda(\E)\gtrsim 0$ at the energy scale $\E$. We cannot determine the value of $\tilde Z_{H}(\E)$, due to unknown $Z^{^S}_{H}(\E)$ and $Z_{HY}$ in Eq.~(\ref{renc}). At the end, we have parameters $\E$, $\tilde Z_{H}(\E)$ and $\tilde \lambda(\E)\gtrsim 0$. With these considerations, we redo the BHL-analysis \cite{bhl1990} of renormalization group equations for the top-quark and Higgs boson masses in the symmetry-breaking phase.

\noindent
{\bf The top-quark and Higgs boson masses.}
\hskip0.1cm 
In the Standard Model of particle physics, using the full one-loop $\beta$-functions (neglect light-quark masses and mixings), the renomalization-group equations for running couplings $\bar g_t(\mu^2)$ and $\bar \lambda(\mu^2)$ are
\begin{eqnarray}
16\pi^2\frac{d\bar g_t}{dt} &=&\left(\frac{9}{2}\bar g_t^2-8 \bar g^2_3 - \frac{9}{4}\bar g^2_2 -\frac{17}{12}\bar g^2_1 \right)\bar g_t,
\label{reg1}\\
16\pi^2\frac{d\bar \lambda}{dt} &=&12\left[\bar\lambda^2+(\bar g_t^2-A)\bar\lambda + B -\bar g^4_t \right],\label{reg2}
\end{eqnarray}
where 
\begin{eqnarray}
A &=&\frac{1}{4}\bar g_1^2+ \frac{3}{4}\bar g^2_2,\quad B=\frac{1}{16}\bar g^4_1 + \frac{1}{8}\bar g^2_1\bar g^2_2 + \frac{3}{16}\bar g^4_2;
\label{reg2ab}
\end{eqnarray}
and, for running gauge couplings of $SU_c(3)$, $SU_L(2)$ and $U_Y(1)$ are
\begin{eqnarray}
16\pi^2\frac{d\bar g_i}{dt}=-c_i\bar g_i^3,
\label{regg}
\end{eqnarray}
with
\begin{eqnarray}
c_1=-\frac{1}{6}-\frac{20}{9}N_g,\quad c_2=\frac{43}{6}-\frac{4}{3}N_g,\quad c_3=11-\frac{4}{3}N_g,
\label{gci}
\end{eqnarray}
where $N_g=3$ is the number of fermion families and $t=\ln\mu$.
Adopting $M_z\approx 91.2$GeV, $M_w\approx 80.4$GeV, $v\approx 239.5$GeV, gauge couplings $\bar g^2_1(M_z)\approx 0.13$, $\bar g^2_2(M_z)\approx 0.45$ and $\bar g^2_3(M_z)\approx 1.5$,
we use the mass-shell condition to
determine the top-quark mass and the Higgs boson mass 
\begin{eqnarray}
m_t=\bar g_t(m_t^2)v/\sqrt{2},\quad m_{_H}^2/2=\tilde \lambda (m_{_H}) v^2,
\label{thmass}
\end{eqnarray}
provided the boundary conditions $\tilde Z_H(\E)$ and $\tilde  \lambda(\E)$ of Eqs.~(\ref{boun0}),(\ref{boun2}) at the energy threshold $\E$ are given.

\begin{figure}
\begin{center}
\includegraphics[height=1.25in]{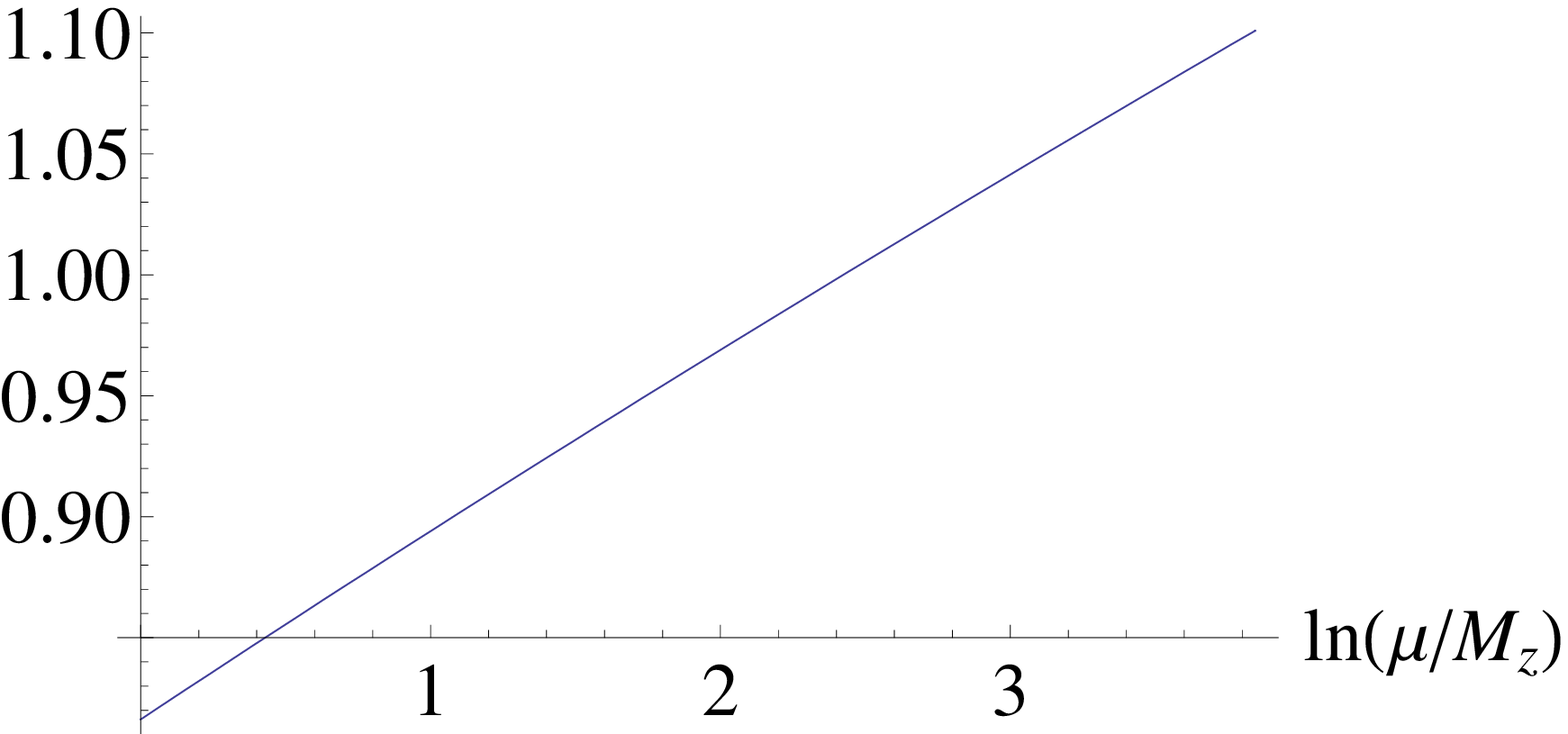}
\includegraphics[height=1.25in]{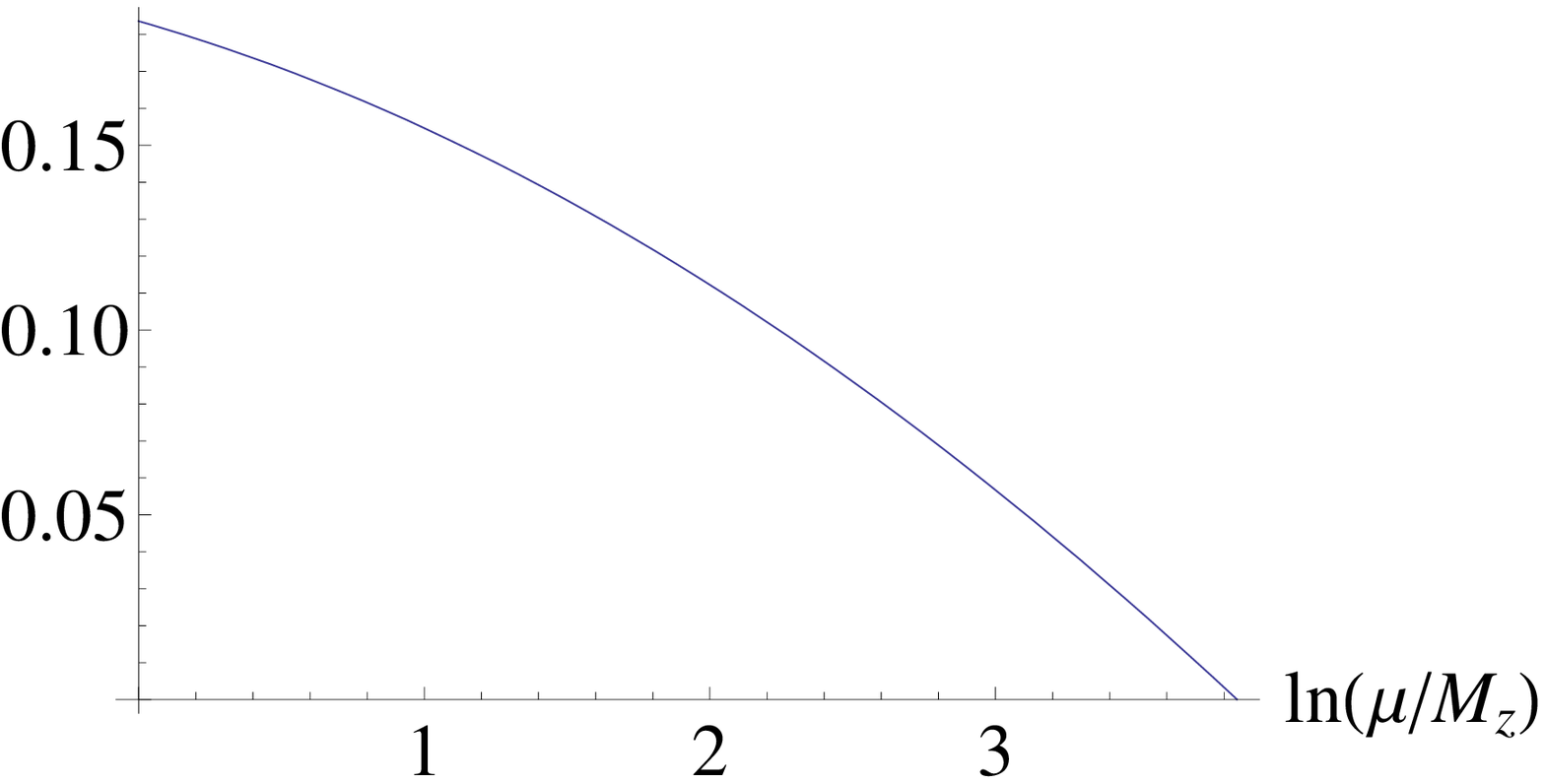}
\put(-370,95){\footnotesize $\tilde Z_{H}(\mu)$}
\put(-170,95){\footnotesize $\tilde \lambda(\mu)$}
\caption{As numerical solutions to the renormalization equations (\ref{reg1}-\ref{gci}), the $\tilde Z_{H}(\mu)$ and $\tilde  \lambda(\mu)$ of Eq.~(\ref{boun0}) are plotted as functions of the running energy scale $\mu$ from the energy threshold $\E\approx 4.27\cdot 10^3 $GeV to the top-quark and Higgs boson mass scales, with $\tilde Z_{H}(\E)\approx 1.101$ and $\tilde  \lambda(\E)\lesssim 10^{-5}$.} \label{figrg}
\end{center}
\end{figure}

The system of Eqs.~(\ref{reg1})-(\ref{thmass}) and boundary conditions is completely determined, provided the boundary values of $\E$, $\tilde Z_{H}(\E)$ and $\tilde  \lambda(\E)$ are given. BHL gave an elegant analytical analysis of fix points of this system and numerical results. Readers are suggested to the original article. Numerically integrating Eqs.~(\ref{reg1})-(\ref{thmass}), we reproduce the BHL result (Fig.~4 and Table.~I in Ref.~\cite{bhl1990}) with $\tilde Z_{H}(\E)=0$ and $\tilde  \lambda(\E)=0$ for selected values of the energy scale $\E$, so as to check our numerical calculations. The question is then the following. In the scenario $\E<\Lambda$, $\tilde Z_{H}(\E)\not=0$ and $\tilde  \lambda(\E)\gtrsim 0$  presented in this Letter, whether or not there is a physically sensible solution to the system of renormalization-group Eqs.~(\ref{reg1})-(\ref{thmass}) and boundary conditions, corresponding to the known top-quark and Higgs boson masses nowadays. In other words, whether or not there are sensible values of $\E<\Lambda$, $\tilde Z_{H}(\E)\not=0$ and $\tilde  \lambda(\E)\gtrsim 0$, for which the system of renormalization-group Eqs.~(\ref{reg1})-(\ref{thmass}) and boundary conditions gives rise to the experimental values $m_t\approx 172.9\,$GeV and $m_H\approx 126\,$GeV. Indeed, our numerical calculations show a numerical solution 
\begin{eqnarray}
\E\approx 4.27\cdot 10^3 {\rm GeV},\quad \tilde Z_{H}(\E)\approx 1.101,\quad {\rm and}\quad \tilde  \lambda(\E)\lesssim 10^{-5}, 
\label{solution}
\end{eqnarray}
which gives rise the $m_t\approx 172.7$GeV and $m_{_H}\approx 126.7$GeV. This solution is sensitive to the values of $\E\approx 4.27\cdot 10^3 $GeV, $\tilde Z_{H}(\E)\approx 1.101$, and insensitive to the value $\tilde  \lambda(\E)\lesssim 10^{-5}$. A small deviation of $\E$ and $\tilde Z_{H}(\E)$ values from the solution (\ref{solution}) results in the deviation of the top-quark and Higgs boson masses from their experimental values. In addition, in the parameter space of $\E$, $\tilde Z_{H}(\E)$ and $\tilde  \lambda(\E)$, there is no another point (solution) to give experimental values of the top-quark and Higgs mass, satisfying the renormalization-group Eqs.~(\ref{reg1})-(\ref{thmass}) and boundary conditions. Corresponding to the numerical solution (\ref{solution}) we found, the renormalization-group evolutions of $\tilde Z_{H}(\mu)$ and $\tilde  \lambda(\mu)$ of Eq.~(\ref{boun0}) are plotted in Fig.~(\ref{figrg}). 

The determined energy threshold $\E$ value, which is about 18 times larger than the electroweak breaking scale $v$, has some physical consequences. The quadratic divergence $\Lambda^2$ in the gap-equation (\ref{delta}) is replaced by $\E^2\ll \Lambda^2$ 
\begin{eqnarray}
\frac{1}{G_c}-\frac{1}{G}=\frac{1}{G_c}\left(\frac{m_t}{\E}\right)^2
\ln \left(\frac{\E}{m_t}\right)^2>0,
\label{delta1}
\end{eqnarray}
where $G_c\approx 8\pi^2/N_c\E^2$.
The unnatural fine-tuning problem is greatly soften by setting the four-fermion coupling
$G/G_c=1+ {\mathcal O}(m_t^2/\E^2)$ and $m_t^2/\E^2\approx 1.64\times 10^{-3}$, instead of the drastically fine-tuning the four-fermion coupling, $G/G_c=1+ {\mathcal O}(m_t^2/\Lambda^2)$ for $\Lambda \gg m_t$. In this case, one can have the physically sensible formula that connects the pseudoscalar (coupling to the longitudinal $W$ and $Z$) decay constant $f_\pi$ to the top-quark mass (see \cite{bhl1990}):
\begin{equation}
f_\pi^2=\frac{1}{4\sqrt{2}G_F}\approx \frac{N_c}{32\pi^2}m_t^2\ln \frac{\E^2}{m_t^2}=\frac{N_c}{32\pi^2}\E^2\left(1-\frac{G_c}{G}\right),
\label{decay}
\end{equation}
without a drastic fine-tuning, where $G_F=1/\sqrt{2}v^2$ is the Fermi constant.

\noindent
{\bf Some remarks.}
\hskip0.1cm 
We are not able to non-perturbatively calculate the energy threshold $\E$, the renormalization-wave function $Z^{^S}_H$ and quartic vertex-coupling $\lambda^{^S}_H$ as functions of energy-momentum, as well as their scaling laws in the neighborhood of the second order phase transition at $G_{\rm crit}$. Nevertheless, the $\E$, $\tilde Z_H(\E)$ and $\tilde\lambda(\E)$ values are completely determined by the self-consistency within the presented theoretical framework in agreement with experimental values of the top-quark and Higgs boson masses, and there is no any free parameter in this determination. 

The energy threshold $\E\approx 4.27\cdot 10^3 \,{\rm GeV}=4.27\,$TeV is the same order of the maximum energy currently reached by the LHC \cite{lhc}. It becomes a chance and challenge to verify whether
the theoretical framework we discussed is true or not.    
The spectra of composite Dirac fermions and their vector-like couplings to intermediate gauge bosons can be possibly examined by checked by measuring the left-right asymmetry \cite{xue_parity}
\begin{eqnarray}
A_{LR}&=&\frac{\sigma_L-\sigma_R}{\sigma_L+\sigma_R}
\label{aslr}
\end{eqnarray}
where $\sigma_L$ ($\sigma_R$) is the cross-section of high-energy ($>\E$) particle colliding with left-handed (right-handed) polarized particles. The signal $A_{LR}\rightarrow 0$ indicates the restoration of 
the parity-symmetry.


To end our Letter, we present a brief discussion what is the possible dynamics at high-energy scale for the origin of effective high-dimensional operators of all fermions fields. Usually composite models for top-quark and Higgs scalar are based on an extended gauge group (strong technicolor) at a higher scale (see for example Ref.~\cite{hill1994}). What is a possible completion of the theory in this Letter at an even higher scale?  We present, on the basis of our previous works on this issue, a brief discussion on the origin of high-dimensional operators of all fermion fields due to the quantum gravity at the Planck length 
($a_{\rm pl}\sim 10^{-33}\,$cm, $\Lambda_{\rm pl}=\pi/a_{\rm pl}\sim 10^{19}\,$GeV).
Studying the quantum Einstein-Cartan theory in the framework of Regge calculus \cite{wheeler1964,regge61}, we recently calculated this minimal length $a\approx 1.2\,a_{\rm pl}$ \cite{xue2010}.
This discrete space-time provides a natural regulator for local quantum field theories of particles and gauge interactions. Based on low-energy observations of parity violation, the Lagrangian of Standard Model was built in such a way as to preserve the exact chiral gauge symmetries $SU_L(2)\otimes U_Y(1)$ that are accommodated by left-handed fermion doublets and right-handed fermion singles. However, a profound result, in the form of a generic no-go theorem \cite{nn1981,nn1991}, tells us that there is no consistent way to straightforwardly transpose on a discrete space-time the bilinear fermion Lagrangian of the continuum theory in such a way as to preserve the chiral gauge symmetries exactly, one is led to consider at least quadrilinear fermion interactions to preserve the chiral gauge symmetries. For
example, the four-fermion operator in the Einstein-Cartan theory can be obtained by integrating over static torsion fields at the Planck scale \cite{torsion}. The very-small-scale structure of space-time and high-dimensional operators of fermion interactions must be very complex as functions of the space-time spacing $\tilde a$ and the gravitational gauge-coupling $g_{\rm grav}$ between fermions and quantum gravity at the Planck scale. We are bound to find an ultra-violet fix point of the gravitational gauge-coupling \cite{xue2012}. As the running gravitational gauge-coupling $g_{\rm grav}(\tilde a)$ is approaching to its ultra-violet critical point $g^{\rm crit}_{\rm grav}$ for $\tilde a \rightarrow a_{\rm pl}$, physical scale $\Lambda = \xi^{-1}[g_{\rm grav}(\tilde a),\tilde a] \ll \tilde a^{-1}$
should satisfy the renormalization group invariant equation in the neighborhood of the ultra-violet fix
point, where the irrelevant high-dimensional operators of fermion interactions are suppressed at least by  
${\mathcal O}(\Lambda/\Lambda_{\rm pl})$; only the relevant operators receives anomalous
dimensions and become renormalizable dimension-4 operators at the scale $\Lambda$ and their effective couplings is larger than the critical value (\ref{gc}).
This is a complicate and difficult issue and needs non-perturbative calculations to show such scaling phenomenon.


\begin{thebibliography}{99}
\bibitem{neutral}
$SU_L(2)$-gauge singlet $\Psi_D=(\Psi_L,t_R)$ can be candidates for cold dark matter, for being very massive and very weak interacting \cite{xue_parity}.
\bibitem{pole-cut}
Bound states of three fermions are presented as poles of four-fermion interacting $S$-matrix, locating at $E_{\rm bind}[G(a),a]<0$ in the energy-momentum plane. When the coupling $a^2G(a)$ becomes smaller, $E_{\rm bind}\rightarrow 0^-$, these poles of the four-fermion interacting $S$-matrix become cuts, representing three elementary fermion states or a composite boson and elementary fermion states.
\end{thebibliography}
\end{document}